\documentclass[twocolumn]{aastex62}

\shorttitle{Cas A Element Abundances}
\shortauthors{Laming et al.}

\begin{document}

\title{Element Abundances in the Unshocked Ejecta of Cassiopeia A}

\correspondingauthor{J. Martin Laming}\email{laming@nrl.navy.mil}

\author[0000-0002-3362-7040]{J. Martin Laming}
\affiliation{Space Science Division, Code 7684, Naval Research Laboratory,
Washington DC 20375, USA}

\author[0000-0001-7380-3144]{Tea Temim}
\affiliation{Space Telescope Science Institute, 3700 San Martin Dr., Baltimore MD 21218, USA}


\begin{abstract}
We analyze and model the infrared spectrum of the Cassiopeia A supernova remnant, with the aim of determining the masses
of various elements in the unshocked ejecta. In this way we complement the survey of the X-ray emitting ejecta of \citet{hwang12} to 
provide a complete census of the elemental composition of the Cas A ejecta. We calculate photoionization-recombination equilibria to determine
the ionization balance of various elements in the ejecta as a function of density, using the X-ray and UV emission from the forward and
reverse shocks as the ionizing radiation. With the assumption that all emission lines are principally excited at the ejecta density that maximizes
their emission, we can convert observed line intensities into element masses. We find that the majority of the $\sim 3 $M$_{\sun}$ ejecta have already been through the reverse shock and are seen today in X-rays.
A minority, $\sim 0.47\pm 0.05$ M$_{\sun}$, with uncertainties quoted here coming from the data fitting procedure only, are still expanding inside the reverse shock and emitting in the infrared. This component is comprised mainly of O, Si, and S, with no Fe detectable. Incorporating uncertainties estimated to come from our modeling, we quote
$0.47 \pm {0.47\atop 0.24}$ M$_{\sun}$.
We speculate that up to a further 0.07 $M_{\sun}$ of Fe may be present in diffuse gas in the inner ejecta, depending on the Fe charge state.

\end{abstract}


\section{Introduction}
Supernovae (SNe) have long been recognized as one of the primary sites for the synthesis of the chemical elements \citep[e.g.][]{arnett96}. With this background,
a long held goal of research in supernova remnants (SNRs) has been to quantify the elemental composition of SN ejecta to investigate
the different sites of nuclear burning and evaluate the various contributions to nucleosynthesis. Such work with the X-ray emitting
reverse shocked ejecta has long been stymied by uncertainties in the shock physics and the non-equilibrium nature of the
plasma, and the variations in such details along the line of sight. \citet{hwang12}, studying the Chandra VLP (1 Million Second) observation of the Cassiopeia~A SNR, were able 
in large part to circumvent many of these problems by taking advantage of the SNR morphology for expansion into a remnant stellar wind, exploiting the high statistical quality of the data, all aided by the current evolutionary state of the SNR that facilitates such an approach. They estimate the reverse shock to be
encountering approximately the innermost 10\% of the ejecta, meaning that the X-ray data yield element abundances for the outer 90\%.
This composition was found to be dominated by O, with smaller amounts of Ne, Mg, Si, S, Ar, and Fe also present. In fact relative
to O, somewhat surprisingly, all these elements are subsolar in abundance, suggesting that much of the original heavy element composition
became locked inside the compact central object. That Cas A was produced by an asymmetric explosion is by now
well-established, by evidence ranging from the recoil of the compact central object \citep{thorstensen01} and directional differences in ejecta velocities from
light echo observations \citep{rest11}, to the bipolar
structure with jet-like features seen in X-ray emitting Si ejecta \citep{hwang04, vink04, laming06}, as well as optical 
\citep[][and references therein]{fesen01} and infrared (IR) emission \citep{hines04}. More recently, \citet{sato20} detect Mn in Cas A and infer the Mn/Cr abundance ratio with implications for the progenitor and evolutionary route
of Cas A to a Type IIb SN, and the explosion energy and asymmetry. \citet{hirai20} give further exploration of such an evolutionary scenario.

In many ways, Fe is the key element in understanding core-collapse explosions and their asymmetries, as it originates at the explosion center,
closest to the mass cut.\footnote{The division between material that falls back onto the compact central object, and material thrown outwards as ejecta.} It is often found together with other intermediate
mass elements (Si, S, Ar, Ca, etc) as a result of incomplete Si burning. Closer to the explosion
center, complete Si burning is expected to produce essentially pure Fe. So too is a burning
regime known as ``$\alpha$-rich freeze out'' (described more fully below) which also produces
$^{44}$Ti. The detection of significant Fe in X-rays in the outer parts of the remnant \citep{hwang09,hwang12}, and the 
NuSTAR location of $^{44}$Ti near the remnant center \citep{grefenstette17} poses a problem that motivates much of the
work in this paper. The detected $^{44}$Ti and Fe are connected to each other as mentioned above
by a common origin in nuclear burning, and both are connected to the neutron star kick by the morphology that
such a kick should imprint on the inner ejecta. We are therefore motivated to investigate the inner unshocked ejecta
in a quantitative manner to study element abundances, to complete a census of the elements produced in the explosion of
Cas~A with particular reference to how much Fe may reside in the inner parts of the remnant, and if possible to study or discuss further any
effects associated with anisotropic nature of the explosion.

The paper is organized as follows. In Section~2 we describe our modeling of the photoionization-recombination (PR) equilibrium used
to calculate the ionization balance of the unshocked ejecta. Section~3 describes the reduction and analysis of the IR Spitzer 
data and Section~4 gives a simple interpretation of the data in terms of element masses in the inner ejecta, while Section~5 gives
a similar analysis of IR emission from ejecta encountering the reverse shock. The emission analyzed here dominates the IR spectrum,
but only accounts for a small fraction of the inner ejecta mass. The emission from the PR equilibrium ejecta is much weaker in terms of IR
signal strength, but accounts for a much larger fraction of the ejecta we wish to study. Section~6 discusses many of the factors that might
complicate our analysis; gas cooling by coupling to dust, gas heating by radioactivity, molecules, and finally limits on the mass of Fe present
inside the reverse shock and consequences this may have for explosion mechanisms, before Section~7 concludes.

\section{Photoionization-Recombination Models}
We calculate the pre-reverse shock ejecta ionization state following initial considerations by \citet{hamilton84}.
The unshocked inner ejecta are in ballistic expansion from the explosion center. They are optically thin to ionizing radiation, and ionization and recombination rates
are sufficiently rapid that the ionization balance may be treated in a PR equilibrium, with the photoionizing radiation being UV to X-radiation from the forward and reverse shocked plasma. A reverse shock radius of 1.7 pc \citep[assuming the 3 M$_{\sun}$ model of][]{hwang12} gives a volume for the inner ejecta of
$6\times 10^{56}$ cm$^3$. Filling this uniformly with 0.3 M$_{\sun}$ of ejecta gives a density
of 0.6 amu cm$^{-3}$, or 0.04 ions cm$^{-3}$ if assumed pure O. The corresponding radial optical depth to X-rays assuming an absorption cross section of order $10^{-22}$ cm$^2$ is then $\tau\sim 5\times 10^{18}\times 10^{-22}\times 0.04\sim 2\times 10^{-5}$ and remains $<< 1$ for all reasonable
parameters.
\citet{delaney14} find an {\em electron} density of 
4.2 cm$^{-3}$, higher than implied by this latter density estimate, indicating that the ejecta are clumped. For a range of densities,
the model temperature is determined as in \citet{laming04} by balancing the heating by photoelectrons (Comptonization of free electrons is negligible)
against cooling by radiation and adiabatic expansion. The radiative cooling rate due to line emission
collisionally excited by electrons goes as $n_e ^2$, where $n_e$ is the electron density, whereas the other two processes vary as $n_e$.
Thus lower temperatures are expected in higher density regions. We ignore effects of optical depths in emission lines in the radiative cooling. This
is justified at the end of this section {\it a posteriori}.

Balancing cooling and heating per unit volume,
\begin{eqnarray}
\nonumber & &{3\gamma\over\gamma -1}\left(n_e+\sum _{ions}n_{ion}\right)k_{\rm B}T{v\over r}+\sum _{ions}n_{ion}n_e\Lambda _{ion}\left(T\right)\\
&=& \sum _{ions} n_{ion}\int f_P\left(E_p\right) \sigma _{photo}\left(E_p\right)\left(E_p-E_{thresh}\right)dE_p
\end{eqnarray}
where $n_{ion}$ is the ion density, and electrons and ions are assumed to be at a common temperature $T$. The first term on
the left hand side represents cooling by adiabatic expansion at velocity $v$ and radius $r$
with adiabatic index $\gamma =5/3$. The second term represents radiative
cooling with loss function $\Lambda _{ion}\left(T\right)$ for each ion in the plasma. For cooling by ground term infra-red lines, 
\begin{equation}
 \Lambda _{ion}\left(T\right)= \sum _{lines}{8.63\times 10^{-6}\over g_0}E\Omega {\exp\left(-E/k_{\rm B}T\right)\over\sqrt{T}}
\end{equation}
where $g_0$ is the statistical weight of the ground (i.e. lower) level, $E$ is the energy of the transition, and $\Omega$ is the effective
collision strength for electron impact excitation. We collect the values of $\Omega$ used in this paper in Table 5 in the Appendix. For cooling by electron
impact excitation to higher lying levels, which can be important at the higher temperatures in our models, we use the tabulations
of \citet{summers79}. We also include the radiative cooling by radiative recombination.
\begin{figure}[b!]
\epsscale{1.0}
\plotone{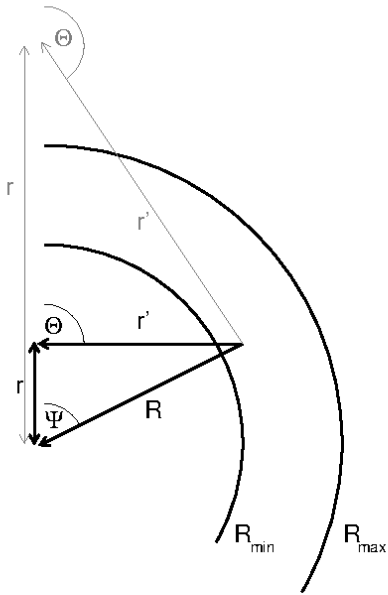}
\caption{Geometry of Cas A SNR for calculation of photoionizing flux. Illuminated points can lie interior (bold lines) or exterior (faint lines) to the shell of X-ray emitting plasma.
The angles $\Theta$ and $\Psi$ refer to the calculation in the Appendix.}
\label{fA1}
\end{figure}
The heating term on the right hand side represents heating by photoelectrons of energy $E_p-E_{thresh}$ where
$E_p$ is the incident photon energy and $E_{thresh}$ is the ionization threshold, integrated over the photon spectrum $f_P\left(E_p\right)$
and the photoionization cross section $\sigma\left(E_P\right)$. 
The flux of photons $J$ seen at radius $r$ in the inner ejecta in terms of the
flux seen by an observer $J_{\infty}$ at radius $r_{\infty}$ is calculated in appendix A and is given by 
\begin{eqnarray}
\nonumber J&=&{3J_{\infty}r_{\infty}^2/2\over R_{max}^3-R_{min}^3}
\{R_{max}-R_{min} \\
\nonumber & &+{R_{max}^2-r^2\over 2r}\ln\left(1+r/R_{max}\over
1-r/R_{max}\right)\\& &-{R_{min}^2-r^2\over 2r}\ln\left(1+r/R_{min}\over
1-r/R_{min}\right)\},
\end{eqnarray}
for the geometry is illustrated in Figure \ref{fA1}. The emitting shell is characterized by maximum and minimum radii $R_{max}$ and
$R_{min}$ and the point at radius $r$ can lie interior (shown with bold lines) or exterior (shown with faint lines), and is discussed in more detail in the Appendix. We take the photoionizing
flux between 300.5 and 9982.9 eV from the {\it Chandra} spectrum \citep[][U. Hwang, private communication 2020]{hwang12}, adding the contribution due to thermal bremsstrahlung only at
each end of this range. This amounts to an X-ray luminosity of $\sim 4\times 10^{37}$ erg s$^{-1}$. The photoionization of lower charge states should be further enhanced by the optical and UV emission coming from ejecta 
density clumps encountering the reverse shock and becoming fast moving knots. However this enhancement is local, the optical and UV does not penetrate
very far into the ejecta before it is absorbed, and as will be seen below, the ejecta of most interest to us are closer to the center, well away from the
reverse shock. Given this ionizing flux, photoionization rates are calculated using the fits to photoionization cross sections of \citet{verner96}. Rates for collisional processes (electron impact ionization, radiative recombination and dielectronic recombination)
are taken from \citet{mazzotta98}, with the exception that many dielectronic rates are been revised and updated in the interim. These newer
references are given in Table 1, with some typographic errors noted.

\begin{figure*}[ht!]
\epsscale{1.0}
\plottwo{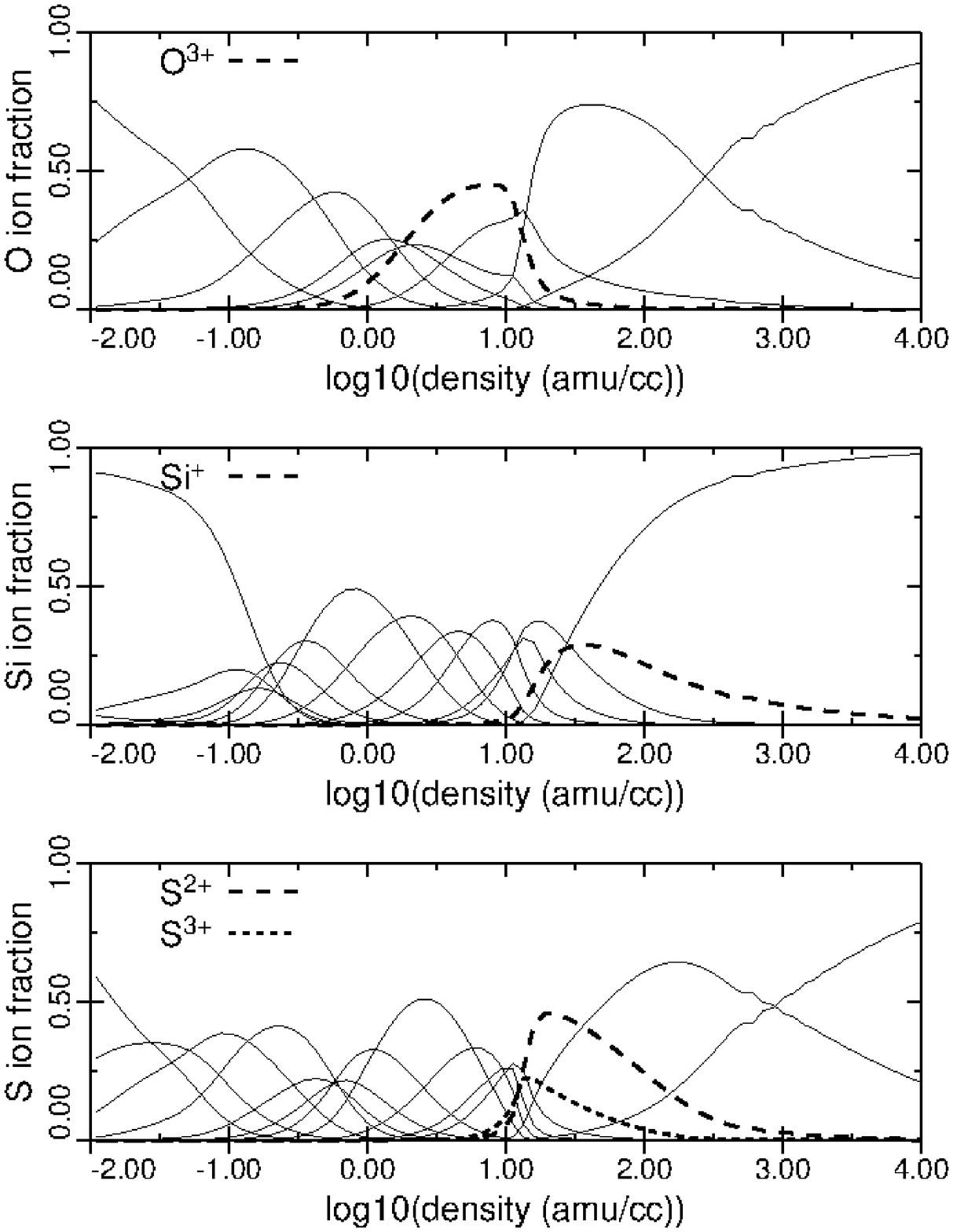}{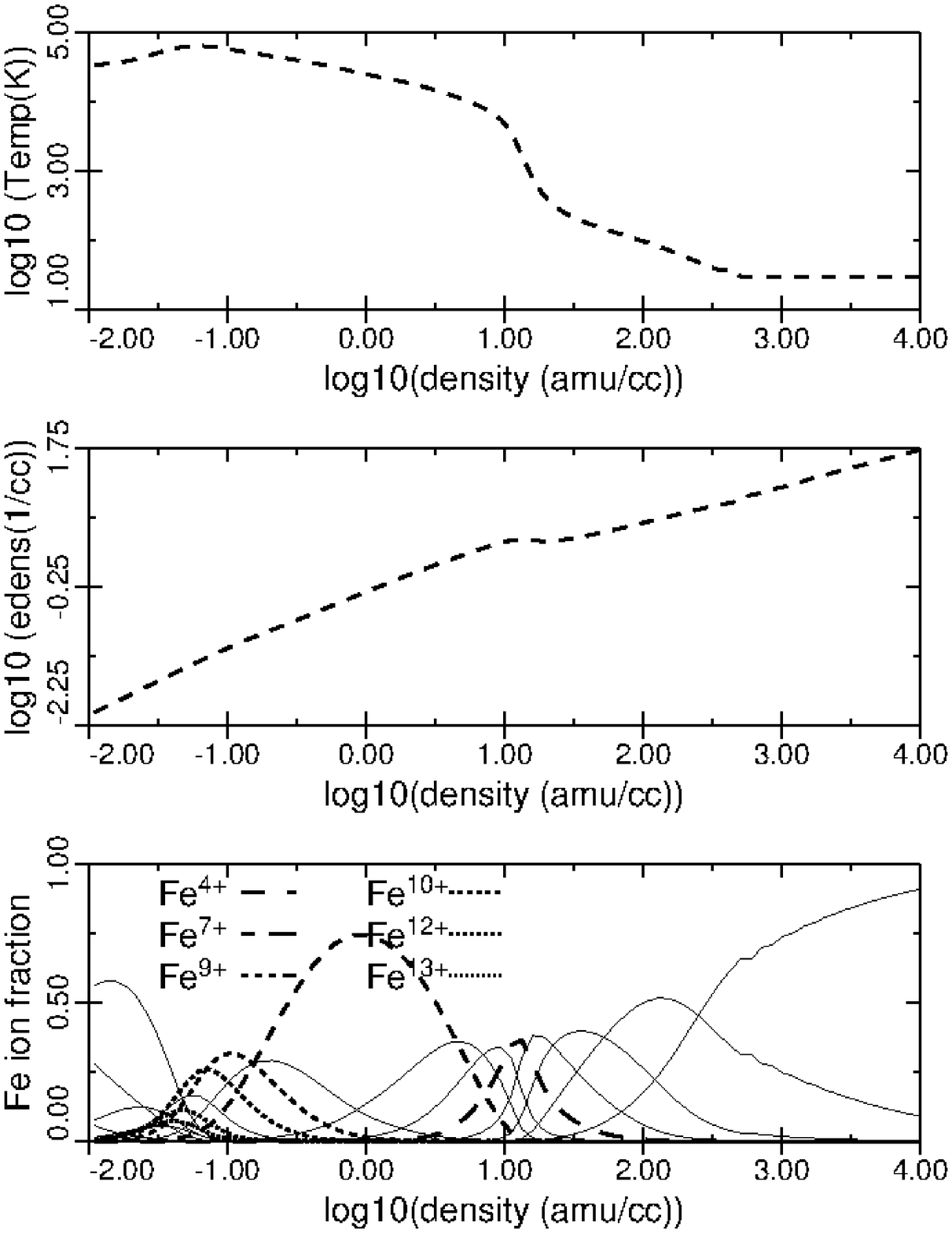}
\caption{Left. Ionization balance of O, Si, and S as a function of ejecta density. Charge states with prominent infra-red
lines are highlighted with thicker lines. Right. Electron temperature, density, and the ionization balance of Fe as a function
of ejecta density, again with important charge states highlighted by thicker lines.}
\label{fejecta}
\end{figure*}

\begin{table}
\begin{center}
\caption{Dielectronic Recombination Data Sources}
\begin{tabular}{ll}
\hline
Isoelectronic Sequence & Reference\\

\hline
H-like & \citet{dasgupta04}\\
He-like& \citet{dasgupta04}\\
Li-like & \citet{colgan04}\\
Be-like & \citet{colgan03}$^a$\\
B-like& \citet{altun04}\\
C-like & \citet{zatsarinny04a}\\
N-like & \citet{mitnik04}$^b$\\
O-like & \citet{zatsarinny03}\\
F-like & \citet{gu03}\\
Ne-like& \citet{zatsarinny04b}\\
Na-like & \citet{gu04}\\
Al-like & \citet{abdel12}\\
Fe 8$^+$ - 12$^+$, 14$^+$ & \citet{badnell06}\\
\hline
\end{tabular}
\end{center}
\tablecomments{$^a$E1 parameter for O$^{4+}$ corrected from 8.552(+2) to 8.552(+3);\\
$^b$E1 parameter for Fe$^{19+}$ corrected from -2.522(+4) to 2.522(+4).}
\end{table}

We calculate the temperature and corresponding ionization balance for a range of densities. We assume that all densities exist within the
reverse shock of Cas A, and that each line will be emitted from the density that maximizes its intensity. Figure \ref{fejecta} shows a sample
run for O- and Si-rich ejecta with trace amounts of S and Fe, chosen to match our eventual result for O, Si, and S (mass fractions O:Si:S:Fe of 
0.3:0.64:0.05:0.01). It is assumed to be a 
distance 0.3 pc from the center of the spherical surface defined by the reverse shock to match ejecta position indicated by the observed Doppler shifts. At the lowest density considered ($10^{-2}$ amu cm$^{-3}$) O is fully ionized. At the average density
quoted above for 0.3 $M_{\sun}$ unshocked ejecta of 0.6 amu cm$^{-3}$, O is principally in the He-like charge state.  As the gas is further
compressed the temperature stays in the range $10^4 - 10^5$ K, until O$^{3+}$ is the dominant O charge state (shown by a thick dashed line
in the top left panel), and strong cooling in the O IV 25.89 $\mu$m
line reduces the temperature to about $10^3$ K. More compression further decreases the average charge state and temperature, until at 
$10^3$ amu cm$^{-3}$, in the absence of any effects of optical depth, the model is no longer able to find an equilibrium and the gas cools unstably. Here we set a minimum temperature of 30 K.
The temperature values are dependent on the element
abundances assumed in the model, and are in good agreement with a temperature of $\sim 100$ K found by \citet{raymond18} from
the ratio of [Si I] 1.645 $\mu$m to [Si II] 34.81 $\mu$m. The lower left panels show the ionization balances
of Si and S under these conditions. On the right hand side we show the temperature, electron density and Fe ionization balance. For the
O, Si and S figures, we have highlighted the charge states with important infra-red emission lines in this work. For Fe, we highlight charge states with important candidate lines for detection by forthcoming instruments (see Table 4, subsection 6.3 below). The charge state fraction shown for Fe$^{7+}$ is possibly spuriously high, due to using the Chandra X-ray spectrum above 300.5 eV, and a pure 
thermal bremsstrahlung spectrum below which reduces its photoionization rate. Although in reality the lower energy radiation will be relatively more absorbed, Fe$^{7+}$ with ionization energy 151.06 eV \citep{kramida19} is more effected by this switch than are Fe$^{8+}$ (ionization energy 233.6 eV) or higher charge states. 

The mass of an element, $M\left(\rm element\right)$, is determined from the unextincted surface brightness of its emission line(s), 
\begin{equation}
J_s ={n_en_{ion}C_{ion}VE\over 4\pi A_{proj}}={n_en_{ion}C_{ion}LE\over 4\pi}
\end{equation}
(in erg cm$^{-2}$s$^{-1}$sr$^{-1}$), by
\begin{eqnarray}
\nonumber M\left(\rm element\right)& =& A_{el}m_pn_{el}V\\ 
&=&{4\pi A_{el}m_pJ_sA_{proj}\over\ n_ef_{ion}C_{ion}E}
\end{eqnarray}
where $A_{el}$ and $n_{el}$ are the atomic mass and number density of the element,
$n_{ion}=f_{ion}n_{el}$ is the ion density in terms of the ionization fraction and 
$n_{el}$, $V=LA_{proj}$ is the SNR volume imaged by the observation, with projected area in the plane of the sky $A_{proj}$,
$m_p$ is the proton mass, and $E$ is the photon energy. We take the denominator $n_ef_{ion}C_{ion}$ to be its maximum value
as calculated as in Fig.2. This is likely to lead to an underestimate of the element mass, though other assumptions below will bias things in the opposite direction. The alternative approach of integrating $f_{ion}C_{ion}$ over
$n_e$ depends on the limits taken for the integration, and is typically a factor of 2 lower than the maximum value of $n_ef_{ion}C_{ion}$, suggesting a factor of 2 higher mass. This would assume that plasma at all densities is present in the inner ejecta of Cas A, though the integral could be weighted by a density profile should one be known.

As mentioned above, extreme clumping may render the emission lines optically thick and reduce the
radiative cooling. We consider the case of [Si II] 34.81 $\mu$m emission. Arguing from the Einstein relations between spontaneous and stimulated
emission/absorption, the absorption cross section is $\sigma = \lambda ^3A_{ji}g_j/g_i\sqrt{2kT/M}= 2\times 10^{-16}$ cm$^2$ for spontaneous
emission rate $A_{ji}=2.17\times 10^{-4}$ s$^{-1}$, wavelength $\lambda = 34.81\times 10^{-4}$ cm, statistical weights $g_j/g_i=2$ and thermal speed
$\sqrt{2kT/M}=2.4\times 10^4$ cm s$^{-1}$. In a knot of radius 1 arc second ($5\times 10^{16}$ cm), the radial optical depth is 0.2 times the product
of ionization fraction, elemental mass fraction and compression over the average ejecta density. At the peak ionization fraction for Si$^+$, this
optical depth evaluates to $\sim 3$ at the line center (taking values from Fig. 2), meaning that a photon will scatter on average 3 times inside the knot before emerging. The electron density from Fig. 2 is $\sim 3$ cm$^{-3}$, which when multiplied by the number of scatterings is still well below the critical 
density of 87 cm$^{-3}$ for this transition, meaning that the photon is not effectively destroyed by the trapping and our assumption of optically thin
emission is valid for the [Si II] emission, and for lines from lower
densities. Such effects may restrict the radiative cooling in higher density regions of the ejecta.

\section{Observations and Data Reduction}

\begin{figure}[t!]
\epsscale{1.0}
\plotone{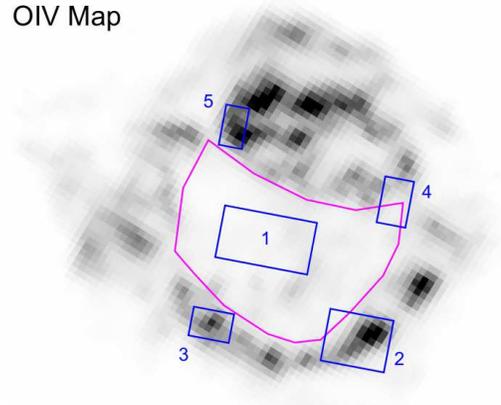}
\caption{Spatial regions observed by the Spitzer IRS (green and magenta) overlaid on the [O IV] 25.89 $\mu$m line map of Cas A. Regions 1-5 are observed at high spectral resolution. We use region 1 for the best sample of inner ejecta. The magenta region shows the extraction area for the low spectral resolution spectra of the inner ejecta.}
\label{fobs}
\end{figure}

Spectroscopy of Cas~A in the 5--37~\micron\ range was obtained with the Infrared Spectrograph \citep[IRS,][]{houck04} aboard the \textit{Spitzer} Space Telescope. The observations were carried out on 2005, Jan 7 (program ID 3310) and 2007, August 29 (program ID 30153). These data consist of a low spectral resolution (R = 57--127) map of the entire SNR \citep{ennis06} and high spectral resolution (R $\sim$ 600) maps of smaller 
regions of Cas A, shown as blue rectangular regions in Figure~\ref{fobs} against the
[O IV] 25.89 $\mu$m emission map generated from data previously published by \citet{isensee12}. North is to the top and east to the left. 
The [O IV] emission is seen to be strongest in the bright ``ring''
surrounding the remnant interior, suggesting that processes connected with the reverse shock dominate the excitation of this line, which are caught in Regions 2-5. Of
most interest to us is Region 1, which encompasses area of 2160 square arcseconds. It is located away from the limb of the reverse shock and more clearly shows
emission from ejecta in PR equilibrium.  The same region is shown in \citet{isensee10} where
it can be seen to encompass a region of strong [Si II] emission. These data were processed with the pipeline version S18.18 and reduced and extracted using the the CUbe Builder for IRS Spectra Maps (CUBISM) software v1.8 \citep{smith07}. A background spectrum from a region adjacent to the SNR was subtracted from the data cube and high and low spectral resolution spectra extracted from the regions shown in Figure~\ref{fobs}. The spectral lines were fit using the Spectroscopic Modeling, Analysis, and Reduction Tool \citep[SMART;][]{higdon04}, the Python \textit{astropy} package \textit{specutils}, as well as with a custom FORTRAN fitting program based on routines in \citet{press86}. The line fitting results  for
Region 1 (of most importance to us) are listed in Tables~\ref{tab2}~and~\ref{tab3}. 

\section{Analysis and Results}
In an initial survey, \citet{eriksen09} considered a single zone model for the inner ejecta, photoionized in the radiation field 
coming from the forward and reverse shocks, treated as a thermal
bremsstrahlung spectrum. In this context, the morphologies of the spectral brightness maps given by \citet{smith09} are
interpreted as follows. The short wavelength emission is excited at the reverse shock where the
electron temperature, $T_e$, is high. In the interior of the remnant, $T_e$ is too low to excite
this radiation. As one moves to longer wavelengths, the images show more center-filled
morphologies, which arise as the excitation thresholds of the transition upper levels reduce to
energies where they can be collisionally excited by thermal electrons with $T_e$ in the range
100 - 500 K. Such a $T_e$ agrees
with that derived from the S III temperature diagnostic, and with our modeling described above.
Further, even the ``center-filled'' images exhibit considerable variability in surface
brightness, indicating inhomogeneous element abundances or density or both. Correspondingly, we concentrate on Region 1. 

\begin{table*}[t!]
\begin{center}
\caption{\label{tab2} High Resolution Ejecta Velocity Components for [O IV] 25.89, [Si II] 34.81, [S III] 33.48 $\mu$m in Region 1.}
\begin{tabular}{lcccccc}
\hline
ion & 1& 2& 3& 4& 5& 6\\

\hline
[O IV] flux (W cm$^{-2}$ sr$^{-1}$)& $2.70\pm 0.02$& $2.11\pm 0.04$& $1.02\pm 0.04$& $2.9\pm 0.2$& $1.91\pm 0.04$& $4.4\pm 0.2$ \\
& $\times 10^{-12}$& $\times 10^{-11}$& $\times 10^{-11}$& $\times 10^{-12}$ & $\times 10^{-11}$& $\times 10^{-12}$\\
LoS velocity (km s$^{-1}$)& -3360& -2074& -270& 1144& 2935& 4344\\
FWHM  (km s$^{-1}$)& 670& 1836& 1462& 883& 3095& 934\\
Implied Mass of O (M$_{\sun}$)& & & $0.027\pm 0.001$& $0.0076\pm 0.0006$& & \\
$f_{PR1}=0.22\pm 0.02$ & & & & & &\\
&&&&&&\\
$\left[{\rm Si~II}\right]$ flux (W cm$^{-2}$ sr$^{-1}$)& & $1.48\pm 0.06$& $2.37\pm 0.14$& 
$5.0\pm 1.2$& $1.63\pm 0.11$& $3.2\pm 1.0$\\
& & $\times 10^{-11}$& $\times 10^{-11}$& 
$\times 10^{-12}$& $\times 10^{-11}$& $\times 10^{-12}$ \\
LoS velocity (km s$^{-1}$)& & -1741& -1542& 862& 2568& 4352\\
FWHM  (km s$^{-1}$)& & 1045& 2344& 1715& 2150& 1310\\
Implied Mass of Si (M$_{\sun}$)& & & $0.067\pm 0.006$ & $0.014\pm 0.003$ & & \\
$f_{PR1}=0.45\pm 0.04$ & & & & & &\\
&&&&&&\\
$\left[{\rm S~III}\right]$ flux (W cm$^{-2}$ sr$^{-1}$)& $1.7\pm 0.5$& $4.80\pm 0.25$& $3.2\pm 0.6$& 
$9.8\pm 2.5$& $5.9\pm 1.3$& $2.4\pm 0.5$ \\
& $\times 10^{-12}$& $\times 10^{-12}$& $\times 10^{-12}$& 
$\times 10^{-13}$& $\times 10^{-13}$& $\times 10^{-12}$ \\
LoS velocity (km s$^{-1}$)& -2357& -1653& -188& 1272& 2321& 2867\\
FWHM  (km s$^{-1}$)& 2212& 1370& 1729& 641& 256& 2321\\
Implied Mass of S (M$_{\sun}$)& &  & $0.0045\pm 0.0007$ & $0.0014\pm 0.0004$ & 
& \\
$f_{PR1}=0.30\pm 0.09$ & & & & & &\\
\hline
\end{tabular}
\end{center}
\tablecomments{Components 1, 2, 5, and 6 are interpreted as coming from the reverse shock, so no mass estimate is given for these. Surface fluxes are not corrected for extinction. Uncertainties in masses come solely from fitting errors. The area of Region 1 is 2160 square arcseconds.}
\end{table*}

\begin{figure}[t!]
\epsscale{1.1}
\plotone{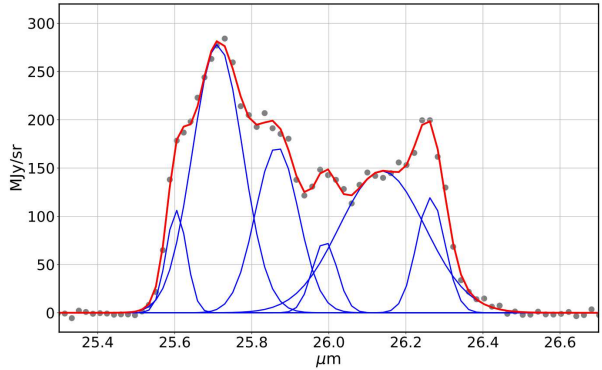}
\caption{The high resolution Spitzer [O IV] 25.89 $\mu$m line profile from Region 1, showing the six component fit given in Table 2; 1 MJy = $4.44\times 10^{-13}$ W cm$^{-2} \mu$m$^{-1}$ at 26 $\mu$m.}
\label{fOIV}
\end{figure}

Figure \ref{fOIV} shows the line profile of [O IV] 25.89 $\mu$m line from Region 1. It is fitted with six Gaussians,
with parameters listed in Table 2. We give the surface brightness, $J_s$,  the Doppler Line of Sight (LoS) velocity, the full width and half maximum (FWHM) of each component and the implied mass of O in
Region 1 assuming PR equilibrium, calculated from
\begin{equation}
M\left(\rm element\right) = {5.87\times 10^{-20}J_s\over f_{ion}C_{ion}n_eE/A_{el}} M_{\sun}
\end{equation}
where the projected area in equation 4 has been replaced by its numerical value, $5.59\times 10^{36}$ cm$^2$, corresponding to an
angular region 54'' $\times $ 40'' at the known distance of 
3.4 kpc.\footnote{Extrapolating to the whole
volume of inner ejecta, the numerical factor $5.87\times 10^{-20}$ would be replaced by $6.05\times 10^{-19}$.} An extinction correction (see below) has also been applied.

Based on the arguments of \citet{eriksen09}, we interpret the [O IV] components 1 and 2, and 5 and 6 as being associated with reverse shock
on the front and back sides respectively. Using the 3 $M_{\sun}$
ejecta mass model of \citet{hwang12}, the ejecta free expansion velocity at the reverse shock should be 4890 km s$^{-1}$, the reverse shock speed
should be 2465 km s$^{-1}$, and therefore the unclumped
post reverse shock decelerated ejecta should have velocity 3042 km s$^{-1}$. These should be taken as ``nominal'' values, being based 
on 1D hydrodynamical models, and considerable variation in the propagation of the reverse shock through the ejecta doubtless exists. Dense clumps experience a 
slower reverse shock, and so are decelerated less and retain a higher fraction
of their free expansion velocity postshock. Additionally,
the Cas A SNR is inferred to be receding from us with velocities given as 859 km s$^{-1}$ \citep{delaney10}, or $760\pm 100$ km s$^{-1}$
\citep{milisavljevic13}, an asymmetry that is also apparent in Table 2. Thus 
components 2 and 5 are reverse shock decelerated unclumped [O IV] emission, while 1 and 6 are [O IV] excited in a shock precursor as in \citet{laming14}, by local photoionization by optical and UV from shocked knots, or represent
the denser less decelerated ejecta clumps going through the reverse shock. The FWHM of components 2 and 5 set upper limits on the reverse shock speed,
indicating 1800 and 3100 km s$^{-1}$ respectively, after subtracting an instrumental width of 500 km s$^{-1}$ \citep{houck04,dasyra08}.
\citet{isensee10} demonstrate that similar velocity profiles hold for [Si II] and [S III], and so we make the same assumption for these lines. This
argument also indicates that the feature at 25.9 $\mu$m is most likely entirely O IV 25.89 $\mu$m, and that [Fe II] 25.98 $\mu$m is hardly present
as a blend in this feature. Pursuing an [Fe II] identification would subtract about 1000 km s$^{-1}$ from the LOS velocities given in table 2 for [O IV]. While there is no reason why components 3 and 4 corresponding to PR equilibrium ejecta should line up, components 1, 2, 5, and 6 representing emission from the reverse shock, or from plasma associated with it should match among the different emission lines. The [O IV] identification is clearly better here than [Fe II], supporting this conclusion. Such an analysis of the various Doppler shifted ejecta plasma components is not possible for Regions 2-5, where the reverse shock and ejecta
expansion velocities are not along our line of sight. Given the increased difficulty
of isolating the PR ejecta in these regions, and the likely smaller contribution that 
it makes to the total emission, we do not consider Regions 2-5 further in this paper.

We infer element masses in region 1 from emission line
intensities using equation 6 assuming an ionization balance computed for a particular set of element abundances, and then iterating our
ionization balance calculations until the element abundances input match those
calculated from the output. We assume a visual extinction $A_v=7$ \citep{delooze17}, K-band extinction
$A_k=0.078A_v=0.546$ \citep{wang19}
and take tabulations and plots (Fig. 8) of $A_{\lambda}/A_k$ from \citet{chiar06} to derive the
extinction at wavelength $\lambda$, $A_{\lambda}$. Consequently we derive $0.035\pm 0.001$ M$_{\sun}$ of O, $0.081\pm 0.005$ M$_{\sun}$ of Si, and $0.006\pm 0.001$ M$_{\sun}$ of S for a total of $0.122\pm 0.005$ M$_{\sun}$  (uncertainties from fits only) in Region 1 from
velocity components 3 and 4, with the remainder of the emission excited by the reverse shock. This mass could be up to a factor of two higher, 
depending on assumptions made in averaging the emission in spectral lines over the range of densities over which they form. The implied reverse shock speed upper limits are
lower for [Si II] and [S III] than they are for [O IV], as expected since these ions are formed in denser regions of ejecta where the reverse shock will
be slower. Also given in Table 2 are values of $f_{PR1}$, the fraction of the total emission in each line coming from plasma in PR equilibrium, i.e. the ratio of components 3 and 4 to the total, for Region 1.

\begin{table*}[ht!]
\begin{center}
\caption{\label{tab3}Cas A Inner Ejecta Element Masses}
\begin{tabular}{lcccccc}
\hline
ion & surface flux (W cm$^{-2}$ sr$^{-1}$) & $n_ef_{ion}C_{ion}E/A_{\rm el}$ (erg s$^{-1}$) & $f_{PR}$ & $10^{0.4A_{\lambda}}$& mass (M$_{\sun}$)\\
\hline
[O IV] 25.89$\mu$m& $3.5\pm 0.2\times 10^{-11}$& $2.81\times 10^{-22}$& $0.15\pm 0.02$& 0.79& $0.14\pm 0.02$ \\
$\left[{\rm Si~II}\right]$ 34.81 $\mu$m& $3.1\pm 0.4\times 10^{-11}$& $2.42\times 10^{-22}$& $0.34\pm 0.03$& 0.86& $0.31\pm 0.05$\\
$\left[{\rm S~III}\right]$ 33.48 $\mu$m& $6.4\pm 1.7\times 10^{-12}$& $4.97\times 10^{-22}$& $0.20\pm 0.08$& 0.84& $0.019\pm 0.009$\\
$\left[{\rm S~III}\right]$ 18.71 $\mu$m& $3.8\pm 0.5\times 10^{-12}$& $1.82\times 10^{-22}$& $0.09\pm 0.06$& 0.75& $0.014\pm 0.010$\\
$\left[{\rm S~IV}\right]$ 10.51 $\mu$m& $5.7\pm 0.5\times 10^{-12}$& $5.99\times 10^{-22}$& $0.18\pm 0.07$& 0.64& $0.016\pm 0.006$\\
$\left[{\rm Ne~II}\right]$ 12.81 $\mu$m& $3.6\pm 0.6\times 10^{-12}$& $2.71\times 10^{-23}$& $\sim 10^{-3}$ & 0.76& $\sim 10^{-3}$\\
$\left[{\rm Ne~III}\right]$ 15.55 $\mu$m& $6.0\pm 0.9\times 10^{-13}$& $1.21\times 10^{-22}$& $0.06\pm 0.04$ & 0.80 & $0.0023\pm 0.0016$\\
$\left[{\rm Ar~II}\right]$ 6.985 $\mu$m& $7.5\pm 0.8\times 10^{-12}$& $6.58\times 10^{-24}$& $\sim 10^{-5}$& 0.70& $\sim 10^{-4}$\\
\hline

\end{tabular}
\end{center}
\tablecomments{Surface fluxes are not corrected for extinction. Uncertainties com from fitting procedures only.}
\end{table*}

\begin{figure}[b!]
\epsscale{1.1}
\plotone{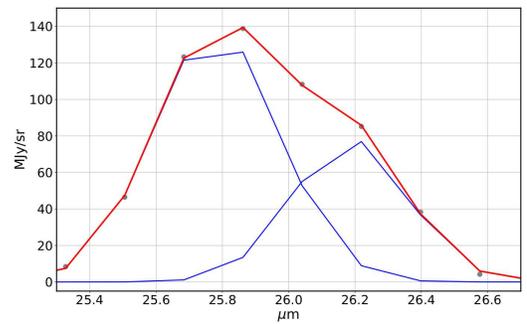}
\caption{The low resolution Spitzer [O IV] 25.89 $\mu$m line profile from the inner magenta contoured region in Fig. 3, showing a two component fit; 1 MJy = $4.44\times 10^{-13}$ W cm$^{-2} \mu$m$^{-1}$ at 26 $\mu$m.}
\label{fOIVl}
\end{figure}

We estimate the masses associated with emission in [O IV], [Si II], and [S III] in the remnant as a whole using surface fluxes measured in the low resolution data in the region enclosed with a magenta line in Fig. 3. An example fit to [O IV] from this region is given in Fig. \ref{fOIVl}.
We need to correct these data 
to isolate the fraction due to PR equilibrium plasma, since at low spectral
resolution this cannot be done by fitting.
Compared to Region 1, the ratio of volume emission to surface emission, i.e. the ratio of PR equilibrium ejecta to that excited by the reverse shock, is lower by a geometrical factor, assuming no other differences. In Region 1, this ratio is just the reverse shock radius, $R_r=1.7$ pc. Taking the spatial extent delineated by the magenta contour to extend out to 84\% of $R_r$, the imaged volume should be $4\pi R_r^3\left(1-\cos ^3\alpha\right)/3$
with surface area of reverse shock $4\pi R_r^2\left(1-\cos\alpha\right)$, where
$\sin\alpha =0.84$. This amounts to a volume to area ratio of 
$0.62 R_r = 1.053$ pc. The fraction of emission from the PR equilibrium plasma
for the inner Region, in $f_{PR}$ in terms of that for Region 1, $f_{PR1}$ is
then $f_{PR}=0.62f_{PR1}/\left(1-0.38f_{PR1}\right)$, and is given in Table 3 for
[O IV], [Si II], and [S III]. This is likely to be an overestimate since it assumes the same density of these emissions throughout the inner ejecta as is seen in Region 1, whereas it appears from the [O IV] and [Si II] morphologies\footnote{Seen most clearly in Fig. 2 of \citet{smith09}.} that Region 1 has more cold dense plasma than elsewhere inside the reverse shock. This effect is difficult to quantify, but is likely to cancel out at least partially the mass underestimate coming from our treatment of the ejecta density above.

We also give masses inferred from
[S III], [Ne II], [Ne III], and [Ar II] measured from the low resolution spectra across the inner magenta region in Fig. 3, extrapolated to the whole region interior to the reverse shock. We estimate $f_{PR}$ for these shorter wavelength lines assuming that the PR emission scales as $\exp\left(-E/k_{\rm B}T\right)$, where $T$ is taken from Fig. 2 for each ion. The values for [Ne III] and [S IV] are very similar to that for [S III], but [Ne II] and [Ar II] are very hard to estimate, so only an order of magnitude is given. Combining the masses of O, Si, S (taken from [S III]), and Ne (from [Ne III]) we derive an inner ejecta mass of $0.47\pm 0.05$ M$_{\sun}$ (uncertainty from fits only), about four times the mass seen in these 
elements in Region 1. Again, an error of up to a factor of two more mass may be present due to our assumption the all lines are radiated from the electron density that maximizes the emission. But in this case another error of similar magnitude but in the opposite direction should be expected due to assumptions surrounding the value of $f_{PR}$, leading us to estimate $0.47 \pm {0.47\atop 0.24}$ M$_{\sun}$ for the mass of the inner ejecta.

\citet{delaney14} give an inner ejecta mass of 0.39 M$_{\sun}$ from free-free absorption at 74 MHz. \citet{arias18} adjust some of the parameters in the analysis of
\citet{delaney14} to find 1.86 M$_{\sun}$, and give their own complete analysis to find $2.95\pm 0.48$ M$_{\sun}$. An important assumption in both works is that the ejecta are assumed homogeneous, with no clumping. This is clearly incorrect, and we are now in a position to quantify this.
The free-free absorption is proportional to a line integral of the square of the electron density, $n_e$, averaged over the Cas A shell,
\begin{eqnarray}
\nonumber\int n_e^2dl& &= {4\over 3}\left(n_{diff}^2\left(1-f^{1/3}\right) + n_{clump}^2f^{1/3}\right)r_{rev}\\
&=& {4\over 3}n_{diff}^2r_{rev}\left(1-f^{1/3}+ x^2f^{1/3}\right),
\end{eqnarray}
where $n_{diff}$ and $n_{clump}$ are the electron densities in the diffuse and clumped ejecta respectively, $f$ is the volume filling factor
for the clumps, $x=n_{clump}/n_{diff}$, and $r_{rev}$ is the reverse shock radius that defines the volume in which the unshocked ejecta reside.
The inferred mass is then
\begin{eqnarray}
\nonumber M&=& {4\pi\over 3}m_p\sum _{el}A_{el}r_{rev}^3\left(n_{el,diff}\left(1-f\right)+n_{el,clump}f\right)\\
&\propto &{1-f+xf\over\sqrt{1-f^{1/3}+x^2f^{1/3}}} \simeq f^{5/6}\quad {\rm for} \quad xf>>1,
\end{eqnarray}
where we have substituted from equation 6 for $n_{diff}$ in the final step, assuming an observed $\int n_e^2dl$. Taking $n_{diff}=\left<n_e\right>/
\left(1-f+xf\right)\simeq 1/30$ as in SN 1987A \citep{li93}, and taking $f=1/10$ we would find an ejecta mass a factor of 0.147 smaller, taking the 
unclumped mass of 2.95 M$_{\sun}$ of \citet{arias18} down to 0.43 M$_{\sun}$. Our inference of unshocked ejecta mass does not depend on such an assumption about the clumping. We simply adopt
the clumping for each ion that maximizes its emissivity, to arrive at an unshocked ejecta mass of order 0.5 M$_{\sun}$, broadly supporting the conclusion of \citet{hwang12} that most of the ejecta has already encountered the reverse shock.

\begin{figure*}[ht!]
\epsscale{1.2}
\plotone{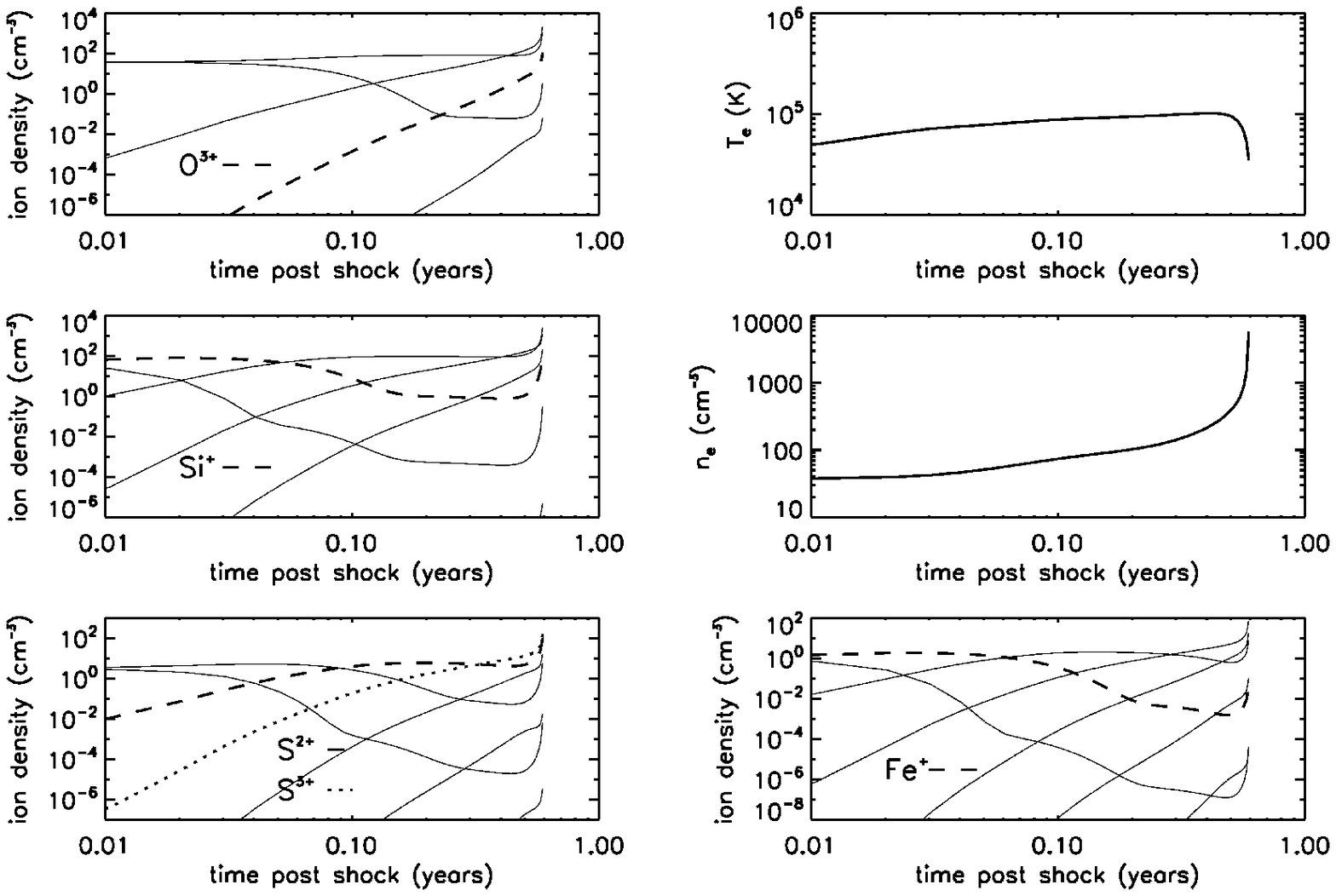}
\caption{Left. Ionization balance of O, Si, and S as a function of time post shock for a knot with overdensity x 1000. 
Charge states with prominent infra-red
lines are highlighted with thicker lines. Right. Electron temperature, density, and the ionization balance of Fe as a function
of time post shock, again with important charge states highlighted by thicker lines.}
\label{knot}
\end{figure*}

\section{Emission from the Reverse Shock}
In this section we give some further justification for our assumptions identifying the velocity components in Fig. \ref{fOIV}. The strongest and most Doppler shifted components were interpreted as coming from the reverse shock location, either from the shock precursor region or from the reverse shocked plasma itself, and therefore only potentially indicative of ejecta composition
at the reverse shock. \citet{docenko10} provide detailed modeling of radiative reverse shocks in ejecta clumps, and give a review of prior
models of this sort. In Fig. \ref{knot} we illustrate the ionization balance for O, Si, S, and Fe, together with the electron temperature and density
as a function of time postshock for the Cas A reverse shock encountering an ejecta clump with a factor of $10^3$ overdensity compared to the
average ejecta density interior to the reverse shock. The composition is O and Si dominated, as for Fig. \ref{fejecta}, where it can be seen that
based on photoionization by X-ray and UV emission from the forward and reverse shocks, these elements are approximately half neutral and half 
singly ionized preshock. In a more realistic calculation, this ionization balance might be further affected by local optical and UV radiation from the
slow reverse shock in the knot itself.

The reverse shock of nominal velocity 2465 km s$^{-1}$ is slowed to about 600 km s$^{-1}$ by the increased density \citep{sgro75}. Upon entering the
knot, the electron temperature, initially $10^4$ K, is increased mainly by collisions with ions to around $10^5$ K, before catastrophic
radiative cooling sets in \citep[following][]{summers79} about 0.6 years ($\sim 2\times 10^7$ s) following shock passage. With these parameters, the
ionizing phase of the shock can account for a [O IV] 25.89 $\mu$m surface brightness of a few $\times 10^{-11}$ W cm$^{-2}$sr$^{-1}$, 
about an order of magnitude too small compared to Fig. \ref{fOIV}, even before the filling factor, which is necessarily $<<1$ is taken into account.
It is evident then that most of the infra-red emission from the reverse shock must come from the radiatively unstable emission setting in about
0.6 years postshock, and this illustrates some of the problems involved in interpreting lines emitted from such an environment in terms of element 
abundances. Not only are we only seeing the plasma at the reverse shock in dense knots that can become radiatively unstable, but this non-equilibrium aspect of the plasma emission renders line emission rates very sensitive to the precise plasma parameters. The temperature and density are changing rapidly in the plasma emitting any particular spectral line, and comparing one spectral line with another formed at different temperature and density will be very uncertain. We argue that the PR plasma still to encounter the reverse
shock is better place to study element abundances in the unshocked ejecta of Cas A, because it is both in an approximate equilibrium and samples a much 
wider distribution of these ejecta.

\citet{docenko10} highlight an important feature ignored in the calculation in Fig. 5, that of the preshock PR region due to local optical and UV-radiation from the reverse shock in the knot itself. This gives rise to a shock precursor with elevated infra-red emission, corresponding to one of our
interpretations of components 1 and 6 in Fig. \ref{fOIV}. This plasma is undecelerated, but is located close to the reverse shock front and back. The reverse shock speed, 2465 km s$^{-1}$, and the post reverse shock ejecta expansion velocity, 3042 km s$^{-1}$, both given above, assume no clumping of the ejecta. Ejecta clumps will encounter a slower reverse shock, and undergo less deceleration, so shocked ejecta should appear at expansion velocities between 3042 and 4890 km s$^{-1}$, on the basis of 1D SNR models. Lower expansion velocities would also be possible if the reverse shock has spent time propagating through underdense ejecta regions, caused e.g. by Fe bubbles inflated by radioactivity early in the evolution of the SNR, causing it to penetrate deeper into the ejecta and encountering less rapidly expanding layers.

\section{Complications}

\subsection{Mass in Molecules and Dust} 
We have adopted the simplest possible model of PR equilibrium to interpret the IR spectrum of Cas A, modeling only neutral atoms and ions. 
\citet{rho09,rho12} detect IR vibrational lines of CO associated with the reverse shock in Cas A, and suggest that such CO must have formed in the year or two following the explosion. If significant quantities of CO formed, less C would then be available to form C-bearing
dust. \citet{wallstrom13} detect high-J rotational lines from CO from a dense knot in Cas A. These authors argue that CO should be dissociated
by the reverse shock, and what is seen is CO reforming in the post shock cooling zone. Estimates of the CO mass in Cas A vary, but are typically in the range $10^{-6} - 10^{-5}$ M$_{\sun}$, and so constitute an insignificant mass compared to that inferred here from the atomic lines.

The formation of dust and molecules following the SN explosion depends sensitively on the ejecta density. \citet{biscaro14} find that significant
clumping of the ejecta in Cas A is necessary for dust formation. Without this, the stripping of the Cas A progenitor by a pre-SN stellar wind renders the ejecta too tenuous and too quickly expanding for molecules and dust to form. \citet{biscaro14} also model the evolution of dust and molecules through the reverse shock. They find that molecules that are destroyed by the shock passage can in general reform in the postshock cooling region, though not to the abundance found in the preshock gas. Dust destroyed by the reverse shock does not reform. \citet{biscaro16} and \citet{micelotta16} consider the post-reverse shock
evolution of dust in more detail, including the destruction of the knots by post-shock instabilities and grain destruction by sputtering, accentuated by
the relative motion between the plasma and the dust grains. The main conclusion is that dust grains are effectively destroyed by these processes at the reverse shock in Cas~A, and considering that the reverse shock has swept through about 90\% of the ejecta \citep{hwang12}, rather little
of the dust originally formed should survive. \citet{bocchio16} perform similar modeling and conclude that most of the dust formed in Cas A should have survived. However, these authors used hydrodynamics solutions suitable for SNR expansion into a uniform interstellar medium (ISM), where the progress of the reverse shock through the ejecta is slower than it would be for a more appropriate assumption of a remnant expansion into a relic stellar wind.

\citet{arendt14} perform a detailed analysis of data from the {\it Spitzer} Infrared Spectrograph (IRS) supplemented by {\it Herschel} Photodetector Array Camera and Spectrometer (PACS) observations at 70, 100, and 160 $\mu$m. After subtracting ISM dust and synchrotron emission they find a mass of warm
(60--130K) dust of $\sim 0.04$ M$_{\sun}$ with components that can be identified spectroscopically as magnesium silicate grains (with low or high Mg to Si ratio) and Al$_2$O$_3$ dust. There are also unidentified featureless components in both the warm dust, and also a larger cold dust
mass comprising up to about 0.1 M$_{\sun}$. \citet{delooze17} analyze a similar dataset further supplemented at long wavelengths by the
{\it Herschel} Spectral and Photometric Imaging Receiver (SPIRE) data, fitting a three part dust model comprising hot (100 - 200 K), warm (40 - 100 K) and cold (10 - 40 K) components. The hot component is found to be consistent with magnesium silicate grains, as in \citet{arendt14}, but the warm and cold components, being relatively featureless, are less well constrained. \citet{delooze17} evaluate the likely dust compositions based on nucleosynthesis
models \citep{nozawa10} for a Type IIb supernova like Cas A, and then evaluate from the observed IR emission limits on the dust mass for each component. Warm and hot dust components comprise $\sim 10^{-3} - 10^{-2}$ M$_{\sun}$, while the cold component can be as high as a few times 0.1 
M$_{\sun}$, considerably more than found by \citet{arendt14}. The main uncertainty seems to be in identifying and correcting for cold dust
emission from the foreground and background ISM. \citet{priestley19} elaborate on this analysis, modeling distributions of dust grains with different sizes and temperatures for each dust component. They assume that the radiation field heating the dust is composed of synchrotron radiation from
the forward and reverse shocks with luminosity $\sim 8\times 10^{36}$ erg s$^{-1}$. By contrast our modeling here takes the full X-ray spectrum
from the forward and reverse shocks detected by {\it Chandra} \citep{hwang12}, not just the synchrotron radiation, with luminosity $4\times 10^{37}$ erg s$^{-1}$. Subsection 4.1.2 of
\citet{priestley19} describes how this increase in heating would decrease the inferred dust mass. A possible ``remedy'' would be to increase the grain size, but  this is unlikely for a Type IIb supernova like Cas A where the progenitor is compact due to pre-supernova mass loss. Hence the post explosion expansion velocities are fast, leading to a rapid decrease in the ejecta density, and inefficient production of large dust grains. Irrespective of grain size, prior work \citep{rho08, barlow10, sibthorpe10} had all given dust masses $< 0.1$ M$_{\sun}$, 
while \citet{dunne09}, interpreting the excess polarization in 850 $\mu$m radiation detected with the Submillimetre Common-Users Bolometer Array (SCUBA) as due to magnetic field aligned dust estimated the cold dust mass in Cas A to be 1.0 M$_{\sun}$. This though is subject to assumptions about the magnetic field direction that might not be justified \citep[see e.g.][]{west17}.

Dust extinctions have also been estimated from the extinction observed from red-shifted ejecta emission from the rear of the SNR compared to blue-shifted lines from the front side \citep{lee15,bevan17}. This last reference infers a dust mass of 1.1 M$_{\sun}$ from the extinction in optical lines of [O I], [O II], and [O III]. However, the ejecta properties assumed in the code for late-time supernovae may not be entirely appropriate for Cas A.
The optical emission from [O I], [O II] and [O III], if excited collisionally, must come solely from dense clumps at the reverse shock. If excited by recombination, the morphology in the interior should resemble that of the cold dense gas, illustrated for [S III]  in \citet{milisavljevic15} or by [O IV], [Si II] or
[S III] in \citet{smith09}. Neither case matches the 
spherical shell of thickness of order 1 pc as modeled in \citet{bevan17}. Additionally, in the case of Cas A, the IR lines of O (see e.g. Fig. \ref{fOIV}) show similar blue-red asymmetry to the optical lines, but should be subject to much less scattering, further implying that the optical depth effects due to dust in Cas A have been overestimated.

In this work we strongly favor the lower dust masses, $< 0.1$ M$_{\sun}$ for the reason advanced by \citet{arendt14}; dust masses approaching 1 M$_{\sun}$ imply an unrealistically high dust condensation efficiency. There is tension between our low estimate of the unshocked ejecta mass, a 
few tenths of a solar mass, and the higher values of the inferred dust mass reported in the literature \citep[e.g.][]{dunne09,priestley19,bevan17,delooze17}. Further investigation is warranted to reconcile gas mass estimates based on the equations of hydrodynamics and morphology \citep{hwang12}, interpretations of the IR line spectrum emitted by atoms and ions in PR equilibrium (this work), and estimates of the dust mass based on modeling of dust compositions, sizes, heating, radiation and assumed locations within the remnant.

\subsection{Cooling by Dust} 
The cooling by dust in Cas A amounts to a luminosity of $10^{36} - 10^{37}$ erg s$^{-1}$, depending on the interpretation given to dust signals
in terms of intrinsic SNR emission versus foreground and background ISM emission. The higher literature dust masses make the corresponding luminosity 
comparable with the X-ray emission from forward and reverse shocked gas. 
Our modeling of the ionization fraction above accounted for heating by this X-ray mission balanced only by cooling by radiation in atomic and ionic emission lines and adiabatic expansion of the gas. No gas coupling to the dust was accounted for, and here we give some justification. \citet{dwek87} gives the gas cooling rate due to 
collisions with dust particles (assumed much colder than the gas) as approximately
\begin{equation}
\Lambda = 10^{-23}\left(T/10^5 ~{\rm K}\right)^{3/2}~ {\rm erg~ cm}^3~ {\rm s}^{-1},
\end{equation}
so that the cooling per unit volume is $\Lambda n_e n_{dust}$ erg cm$^{-3}$ s$^{-1}$. The electron density, $n_e$ can be taken from Fig.
\ref{fejecta} and the average dust grain density is
\begin{equation}
\left<n_{dust}\right>=\left(8\times 10^{-7}\over \rho _{dust}\right)\left(M_{dust}\over M_{\sun}\right)\left(r_{dust}\over 0.01~\mu{\rm m}\right)^{-3}
\end{equation}
where $\rho _{dust}\sim 1$ g cm$^{-3}$ is the dust grain mass density, $M_{dust}$ is the dust mass contained within the reverse shock 
and $r_{dust}$ is the dust grain radius. For gas with temperature in the range 100 - 1000 K corresponding to overdensities of $\times 10 
- \times 10^4$ in Fig. \ref{fejecta}, the cooling rate due to dust is in the range $10^{-32} - 10^{-27}$ erg cm$^{-3}$ s$^{-1}$, which is several orders of magnitude lower than the atomic and ionic radiative cooling rates already implemented in the calculation in Fig. 2, and can therefore be neglected. Additionally, the warm or hot dust found in IR studies is also of comparable temperature ($\sim 100$ K) to the gas, further limiting the cooling effect. Dust grains are much less likely in the underdense regions
of ejecta, but even there, the gas cooling rate due to dust is still many orders of magnitude too low to be significant. Dust grains about 0.01 times
the radius above are required, but dust grains of this size are single atoms.

\begin{table*}[ht!]
\begin{center}
\caption{Cas A Fe Coronal Forbidden Lines.}
\begin{tabular}{lccccccc}
\hline
transition and & Fe density& temperature& el. dens.& ion fraction& surface flux& flux at earth& Fe mass\\
wavelength (\AA\ ) &  cm$^{-3}$& K & cm$^{-3}$&  &  cm$^{-2}$ s$^{-1}$ sr$^{-1}$& cm$^{-2}$ s$^{-1}$ & M$_{\sun}$\\
\hline
$\left[{\rm Fe~VIII}\right]$ 54466 & $1.5\times 10^{-2}$& $1.5\times 10^4$&  0.40& 0.74&$3.9\times 10^8$ & 98 & 0.55\\
$\left[{\rm Fe~X}\right]$  6376  & $8.1\times 10^{-4}$& $2.2\times 10^{4}$& 0.058& 0.33& $4.0\times 10^6$ & 1.0 & 0.069\\
$\left[{\rm Fe~XI}\right]$ 7892   & $4.0\times 10^{-4}$& $2.1\times 10^{4}$& 0.036& 0.27& $1.5\times 10^6$& 0.36 & 0.041\\
$\left[{\rm Fe~XIII}\right]$ 10747,10798 & $7.3 \times 10^{-5}$& $1.8\times 10^{4}$& 0.027& 0.073& $3.3\times 10^5$& 0.082 & 0.029\\
$\left[{\rm Fe~XIV}\right]$ 5303 & $2.9 \times 10^{-5}$& $1.6\times 10^{4}$& 0.023& 0.033& $4.0\times 10^4$& 0.012& 0.024\\
\hline

\end{tabular}
\end{center}
\tablecomments{Fe VIII \citet{delzanna14}; Fe X \citet{aggarwal05b}; Fe XI \citet{delzanna10}; Fe XIII \citet{aggarwal05a}; Fe XIV \citet{aggarwal15}. Surface flux is unextincted.}
\end{table*}

\subsection{Radioactivity and Estimates for Fe} 
The X-ray survey of Cas A ejecta of \citet{hwang12} found up to 0.13 M$_{\sun}$ of Fe ejecta, approximately the total amount expected to have been
ejected in the explosion \citep{eriksen09b}. The fact that all or most of this Fe is found in the outer layers of ejecta is a surprise, and a major motivation for studying the inner
unshocked ejecta is to place limits on the Fe that can be found there. The inner ejecta are clumped, most likely by the expansion of Fe-Co-Ni
bubbles by their radioactivity \citep[e.g.][]{li93}. Thus most of the Fe is expected to be in the tenuous gas surrounding the more strongly
radiating clumps from which the lines we study here are emitted. 

The strongest Fe line to be expected from the clumped gas is probably the [Fe II] 25.98 $\mu$m line, maximized when the clumping is around a factor
of $10^2 - 10^3$, according to Fig. \ref{fejecta}. This line is strongly blended with [O IV] 25.89 $\mu$m, and \citet{isensee10} demonstrate that the observed velocity structure in this line compared to the unambiguously identified [Si II] 34.81 $\mu$m and [S III] 33.48 $\mu$m  makes the [O IV]
identification much more plausible. If [Fe II] 25.98 $\mu$m is not detected, the putative identifications of the intrinsically weaker [Fe II] 35.35 $\mu$m
and [Fe II] 17.94 $\mu$m \citep{smith09} in the PR equilibrium ejecta are also doubtful. \citet{isensee12} clearly see the 17.94 $\mu$m line,
and we interpret this as coming from the reverse shock. \citet{gerardy01} see even higher excitation [Fe II] lines with wavelengths in the range
$\sim 1 - 2 \mu$m which are clearly from fast moving knots or quasi-stationary flocculi, and therefore must result from a shock-clump interaction. 

\citet{koo16} and \citet{lee17} present mid and near IR observations of [Fe II] from
ejecta and CSM knots in Cas A. \citet{koo16} demonstate that in radiative shock models
of this emission, most of the [Fe II] radiation comes from the postshock cooling zone, 
similarly to our Fig. 5 above. \citet{lee17} find ejecta knots with abundances commensurate with O burning, with elements like Si and S present, as well as a sample
of ejecta knots showing strong [Fe II] lines but no Si or S. They interpret these
knots as products of complete Si-burning leaving almost ``pure'' Fe, as found in X-rays
by \citet{hwang09,hwang12}, or in some cases the products of $\alpha$-rich freezeout. It is unclear
why such ejecta should be in knots, as opposed to diffuse clouds inflated by radioactivity, but compared to SN 1987A, Cas A was a smaller progenitor at explosion, with presumably
faster expansion and lower opacities leading to a reduced effect of this sort. \citet{koo18}
estimate a mass of [Fe II] emitting shocked dense ejecta of $\sim 3\times 10^{-5}$ M$_{\sun}$. Extrapolating this clumping to the whole inner ejecta is difficult to do with any 
accuracy, but assuming that the Fe clump density is the same throughout the inner ejecta as it is
at the reverse shock, the total Fe mass should be $3\times 10^{-5}$ M$_{\sun}$ multiplied 
by the ratio of the volumes represented by the inner ejecta ($6\times 10^{56}$ cm$^3$) and the reverse shock emission. This last piece is ambiguous, but taking it to be $4\pi R_r^2
\times 2r_{knot}\sim 8\times 10^{55}$ cm$^3$, where $R_r=1.7$ pc $=5\times 10^{18}$cm is
the reverse shock radius and the knot diameter $2r_{knot}\sim 2.5\times 10^{17}$ cm gives
a ratio of volumes of 7.5 and an implied Fe mass in knots of $2\times 10^{-4}$ M$_{\sun}$.
This could be uncertain by at least an order of magnitude either way, but is likely to be 
smaller than the Fe mass in the diffuse Fe-Co-Ni bubbles, estimated below. Features coinciding with
[Fe V] 20.85 and 36.34 $\mu$m, which would imply
Fe masses of order $\sim 10^{-3}$ M$_{\sun}$, are also visible in the inner ejecta region. These only appear at marginal significance and must wait better data for
confirmation.

The [O IV] 25.89 $\mu$m and [Si II] 34.81 $\mu$m morphologies \citep{smith09} and the mass estimate of \citet{arias18} suggest that most of the inner ejecta mass
expected by \citet{hwang12} is clumped into a volume approximately 1/15 - 1/10 of that defined by the interior to the reverse shock. This extra ``space''
is filled by diffuse gas, most likely Fe. Detecting emission from this diffuse ejecta is beyond the sensitivity of the {\it Spitzer} IRS. \citet{li93}
estimate an expansion by a factor of $\times 30$ in SN 1987A, which if applied to Cas A would imply 0.01 M$_{\sun}$ of Fe behind the
reverse shock, assuming the relevant model of \citet{hwang12}. This expansion may well be an overestimate for Cas A, and the corresponding Fe mass an underestimate, because as a smaller
progenitor at explosion with lower opacities, Fe-Co-Ni bubble expansion may well not have been as dramatic. But it remains likely that the Fe mass in these bubbles is greater than
that in the dense knots, which is as it should be if the bubble inflation is to be the 
cause of the knot compression. In Table 4, we give  predicted line fluxes and associated Fe masses for Fe forbidden lines expected in emission from these ejecta. These are calculated assuming an
ejecta composition 65\% Fe and 35\% He by mass, similar to what might be expected in $\alpha$-rich freeze out \citep[e.g][]{arnett96,chieffi17}. As mentioned above, the flux quoted for [Fe VIII] is likely to be an overestimate. A fit to the 5.447
$\mu$m region of the Spitzer low resolution spectrum of the inner ejecta region formally returns a line integrated surface flux of $\sim 10^{-12}$ W cm$^{-2}$ sr$^{-1}$ or in photons $\sim 2.5\times 10^7$ cm$^{-2}$ s$^{-1}$ sr$^{-1}$, which implies a Fe mass of $\sim 10^{-2}$ M$_{\sun}$,
assuming $f_{PR}\simeq 0.2$. Uncertainties are difficult to evaluate, being at the limit of the statistical precision of the data, but clearly the 0.55 M$_{\sun}$ of Fe$^{7+}$ predicted in Table 4 is not supported by the data. Therefore, Fe in the diffuse ejecta must be more highly ionized than Fe$^{7+}$, and we estimate a maximum Fe mass in the diffuse ejecta of 0.07 M$_{\sun}$. We hope that future observations of this sort can places limits on fluxes of some of these other lines to further constrain the diffuse Fe mass in Cas A.

While Cas A is suspected to host up to 0.1 M$_{\sun}$ of dust \citep[e.g.][]{arendt14}, it is unlikely that much Fe can be hidden there. 
Metallic Fe dust in the form of needles has been postulated \citep{dwek04,gomez05} to explain the observed SCUBA submillimetre fluxes at 
450 and 850 $\mu$m. More recently, however, \citet{kimura17} show experimentally that pure Fe grains do not efficiently nucleate; the sticking
probability is very small, meaning that most Fe in grains must be in Fe bearing compounds. Additionally, Cas A was a relatively energetic explosion with a small progenitor meaning that rapid expansion in the early phase of the evolution of the remnant would have inhibited grain formation, especially so for the
radioactively heated Fe ejecta.

We are forced again to the conclusion that the majority of the Fe ejected in the explosion that formed Cas A was placed in the outer ejecta
layers and is now visible in X-rays. However the $^{44}$Ti that would have formed with some of the Fe is almost exclusively within the reverse
shock, with a very different spatial distribution to the Fe \citep{grefenstette17}. In the plane of the sky, the $^{44}$Ti is concentrated on one side 
of the compact central object (CCO), the opposite side to its motion, as would be expected for a hydrodynamic kick where the recoil is taken up by the
ejecta. Further, the ejecta as a whole are also recoiling in this direction \citep{hwang12}, with a velocity of 700-800 km s$^{-1}$ seen especially
in O, Si, S, and Ar. \citet{katsuda18} suggest that such a recoil in intermediate mass elements is a generic feature of neutron star kicks in asymmetric explosions. Using the emission measure as a proxy for element mass, \citet{holland20} infer stronger asymmetries with increasing element mass.
\citet{iyudin19} speculate that the $^{44}$Ti may be trapped in dust grains, since the fluxes of 67.9 and 78.4 keV photons from the nuclear
de-excitation of $^{44}$Sc are marginally lower than the 1157 keV $^{44}$Ca lines, and at lower energy may be absorbed more. The plausibility of the
Ti condensing out of the hot Fe-Co-Ni bubble to form grains while the Fe apparently does not is not discussed, but such a mechanism also offers an
explanation of the different morphologies observed in $^{44}$Ti and Fe. $^{44}$Ti becomes trapped in grains near the remnant center, while the Fe
becomes heated in bubbles and expands into the outer layers of ejecta, leaving the $^{44}$Ti behind. The heating provided by the $^{44}$Ti decay
evaluates to an average of $3\times 10^{-25}$ erg cm$^{-3}$ s$^{-1}$, coming from the Auger electrons emitted by $^{44}$Sc following the K-shell
electron capture decay of $^{44}$Ti. $^{44}$Sc decays to $^{44}$Ca by direct positron emission \citep{hernandez14} followed by annihilation with a
plasma electron to 511 keV photons. If trapped in a grain, the $^{44}$Ti decay will heat the grain, but if free, the energy released amounts to about 1\%
of the heating by photoelectrons, and probably has an insignificant effect on the thermal instability in Fig. 2.

For completeness,  
a final hypothesis to mention is that of \citet{ouyed11}.
The CCO formed in the explosion is suggested to transition (explosively) to a strange-quark star, irradiating the inner ejecta with protons and neutrons.
The resulting spallation reactions break up $^{56}$Ni in favor of lighter elements, $^{44}$Ti and $^{12}$C. While an attractive idea for understanding the
$^{56}$Fe and $^{44}$Ti morphologies, quenching the Fe-Co-Ni radioactivity would surely have significant consequences for the morphologies of other
elements in the inner ejecta. If Fe bubbles cannot form, no obvious mechanism for forming fast moving
knots or other density structures studied here and elsewhere would be available.

\section{Conclusions}
We have derived a mass for the inner ejecta of Cas A of $0.47\pm {0.47\atop 0.24}$ M$_{\sun}$, with the quoted uncertainty dominated by and estimated from model assumptions. A significant innovation in this work lies in the application of a calculation of photoionization-recombination equilibrium for the inner ejecta, which couples the charge state  an element may be ionized to with the surrounding ejecta density, and enables our analysis.
We have assumed that all lines are emitted from the ejecta density where their emission is maximized. This results in an underestimate of the true ejecta mass, but is a simple parameter-free way to proceed. We have measured line intensities using the low resolution {\it Spitzer} data from as much of the inner ejecta as we can while excluding as much as possible emission from the radiative reverse shock. A further correction to exclude reverse shock emission along the line of sight is defined with reference to high spectral resolution data from the brightest internal region (Region 1 in Fig. 3). This likely leads to an overestimate of emission form the PR plasma, again by an amount that is difficult to quantify, but likely to be of the same order as the underestimate coming from our treatment of the ejecta density. Taken together, some degree of cancellation between these two assumptions may be expected. Considerably more elaborate procedures could be applied, but in our view the quality of the {\it Spitzer} data makes such efforts premature.

Our inner ejecta mass estimate, certainly less than 1 M$_{\sun}$, conflicts with measurements of the dust mass also of this order, implying a very high dust condensation efficiency. We significantly favor dust masses of order 0.1 M$_{\sun}$.
We find a composition dominated by O and Si, with a smaller amount of S. Ne and Ar may also be present. Relative to O, there is considerably more Si and S than in the X-ray emitting ejecta studied by \citet{hwang12}, as should probably be expected in the inner ejecta regions. However there is no detection of Fe. Extrapolating from previous work \citep{koo18} we estimate a mass of Fe in knots of $2\times 10^{-4}$ M$_{\sun}$, and an upper limit for Fe in diffuse regions of the ejecta of 0.07 M$_{\sun}$. Thus the conclusion \citep{hwang12} that most of the Fe produced in the explosion is now in the outer ejecta and is visible in X-rays still holds as a challenge to theory.

\acknowledgements This work was supported by a grant from the NASA
Astrophysics Data Analysis Program, NNH16AC24I. JML was also supported by Basic Research
Funds of the Office of Naval Research. We are indebted to Una Hwang for help in the early stages of this project, and acknowledge
the detailed report of the anonymous referee.

\software{CUBISM (v1.8; Smith et al. 2007), SMART (Higdon et al. 2004), astropy (The Astropy Collaboration 2013, 2018)}

\appendix
\section{Illumination and Excitation of the Inner Ejecta}
We model the inner ejecta of Cas A as being in photoionization-recombination equilibrium. The photoionizing
photons come from the UV and X-ray emissions from the forward and reverse shocked plasma. Here we calculate
the flux of photoionizing photons seen by the inner ejecta in terms of the flux seen by an observer at Earth.
Figure \ref{fA1} illustrates the geometry. The photoionized ejecta is at radius $r$, and the emitting ejecta is at 
radius $R$, and confined in a spherical shell between radii $R_{min}$ and $R_{max}$. We choose 
$R_{max}=1.95$ pc, the radius of the contact discontinuity, and $R_{min}=1.91$ pc, representing a surface inward
by 2.5'', the angular radius of the spherical shell into which most of the ejecta are compressed by the hydrodynamics
for a supernova remnant expanding into an exterior stellar wind density profile. The distance between the
photoionized ejecta and the place where the photoionizing flux was emitted is $r^{\prime}$. Figure \ref{fA1} shows
two values of $r$ and $r^{\prime}$, one in regular font illustrating the photoionizing flux received by the inner ejecta,
one in lighter font illustrating the flux received by an observer outside the SNR shell. For an emissivity per unit 
volume of $\epsilon$ the photoionizing flux $J$ is
\begin{equation}
J={\epsilon\over 4\pi}\int {2\pi R^2dR\over r^{\prime 2}}d\left(\cos\psi\right)={\epsilon\over 4\pi}\int{2\pi R dR\over \sqrt{R^2-r^2\sin ^2\theta}-r\cos\theta}
d\left(\cos\theta\right) 
\end{equation}
where $R^2=r^2+r^{\prime 2}+2rr^{\prime}\cos\theta$ and $r^{\prime}\cos\theta +r = R\cos\psi$ 
so that $d\left(\cos\psi\right) = \left(r^{\prime}/R\right)d\left(\cos\theta\right)$. Then
\begin{equation}
J={\epsilon\over 2}\int {r^{\prime}+r\cos\theta\over r^{\prime}}dr^{\prime}d\left(\cos\theta\right)
\end{equation}
which integrates over $r^{\prime}$ between $\sqrt{R_{min}^2-r^2\sin^2\theta}-r\cos\theta$ and 
$\sqrt{R_{max}^2-r^2\sin^2\theta}-r\cos\theta$ to
\begin{equation}
J={\epsilon\over 2}\int_{-1}^{1}\left\{R_{max}\sqrt{1-{r^2\sin ^2\theta\over R_{max}^2}}
-R_{min}\sqrt{1-{r^2\sin ^2\theta\over R_{min}^2}}+r\cos\theta \ln{R_{max}\sqrt{
1-r^2\sin ^2\theta/R_{max}^2}\over R_{min}\sqrt{
1-r^2\sin ^2\theta/R_{min}^2}}\right\}d\left(\cos\theta\right).
\end{equation}
The last term integrates to zero in the interval $-1\le \cos\theta\le 1$. The
two remaining terms are integrated \citep[see][2.271, 3]{gradshteyn65} to give
\begin{equation}
J={\epsilon\over 2}\left\{R_{max}-R_{min} +{R_{max}^2-r^2\over 2r}\ln\left(1+r/R_{max}\over
1-r/R_{max}\right)-{R_{min}^2-r^2\over 2r}\ln\left(1+r/R_{min}\over
1-r/R_{min}\right)\right\}.
\end{equation}
For positions exterior to the SNR shell,
\begin{equation}
J={\epsilon\over 2}\left\{R_{max}-R_{min} +{R_{max}^2-r^2\over 2r}\ln\left(r/R_{max}+1\over
r/R_{max}-1\right)-{R_{min}^2-r^2\over 2r}\ln\left(r/R_{min}+1\over
r/R_{min}-1\right)\right\}.
\end{equation}
As $r\rightarrow\infty$,
\begin{equation}
J_{\infty} ={\epsilon\over 2}{2\over 3r^2}\left(R_{max}^3-R_{min}^3\right)={\epsilon\over 4\pi r^2}{4\pi\over 3}\left(R_{max}^3-R_{min}^3\right)
\end{equation}
so in terms of the unabsorbed flux observed at infinity,
\begin{equation}
J={3J_{\infty}r_{\infty}^2/2\over R_{max}^3-R_{min}^3}
\left\{R_{max}-R_{min} +{R_{max}^2-r^2\over 2r}\ln\left(1+r/R_{max}\over
1-r/R_{max}\right)-{R_{min}^2-r^2\over 2r}\ln\left(1+r/R_{min}\over
1-r/R_{min}\right)\right\}.
\end{equation}
We take $J_{\infty}$ from fits to Chandra observations.

Finally Table 5. collects the values of the collision strengths used to 
evaluate the radiative cooling from the photoionized inner ejecta.

\begin{table}
\begin{center}
\caption{Effective Collision Strengths for IR Lines}
\begin{tabular}{lcccllcccl}
\hline
ion & transition &$\lambda$ ($\mu$m)& $\Omega$& ref& ion & transition &$\lambda$ ($\mu$m)& $\Omega$& ref\\

\hline
O I & 2p$^{4~3}$P$_2$ - 2p$^{4~3}$P$_1$ & 63.09& 0.045& K95& Ar II &2p$^{5~2}$P$_{3/2}$ -  2p$^{5~2}$P$_{1/2}$& 6.985& 2.48& P95\\
& 2p$^{4~3}$P$_2$ - 2p$^{4~3}$P$_0$ & 44.15& 0.015&    &   Ar III&  3p$^{4~3}$P$_2$ - 3p$^{4~3}$P$_1$& 8.991& 3.73& G95\\
O III& 2p$^{2~3}$P$_0$ - 2p$^{2~3}$P$_1$& 88.18& 0.50& L94 & &           3p$^{4~3}$P$_2$ - 3p$^{4~3}$P$_0$& 6.368& 0.70& \\
& 2p$^{2~3}$P$_0$ - 2p$^{2~3}$P$_2$& 32.59& 0.25& &          Ar V& 3p$^{2~3}$P$_0$ - 3p$^{2~3}$P$_1$& 13.07& 4.16& G95\\
O IV & 2p$^{~2}$P$_{1/2}$ - 2p$^{~2}$P$_{3/2}$ & 25.89& 1.61 & A08 & &  3p$^{2~3}$P$_0$ - 3p$^{2~3}$P$_2$& 4.928& 1.87& \\
Ne II& 2p$^{5~2}$P$_{3/2}$ -  2p$^{5~2}$P$_{1/2}$& 12.814& 0.3& W17  & Ar VI& 3p$^{~2}$P$_{1/2}$ - 3p$^{~2}$P$_{3/2}$ & 4.53& 2.95 & S96\\
Ne III&  2p$^{4~3}$P$_2$ - 2p$^{4~3}$P$_1$& 15.555& 0.64& W17&  Ar X&  2p$^{5~2}$P$_{3/2}$ -  2p$^{5~2}$P$_{1/2}$& 0.552& 1.03& S94\\
&  2p$^{4~3}$P$_2$ - 2p$^{4~3}$P$_0$& 10.86& 0.15& & Ar XI&  2p$^{4~3}$P$_2$ - 2p$^{4~3}$P$_1$& 0.692& 0.21& B94 \\
Ne V&  2p$^{2~3}$P$_0$ - 2p$^{2~3}$P$_1$& 14.32& 1.83& L94 & &  2p$^{4~3}$P$_2$ - 2p$^{4~3}$P$_0$& 0.547& 0.048& \\
& 2p$^{2~3}$P$_0$ - 2p$^{2~3}$P$_2$& 8.99& 3.23&  & Ar XIII& 2p$^{2~3}$P$_0$ - 2p$^{2~3}$P$_1$& 1.016& 0.0817&  \\
Ne VI & 2p$^{~2}$P$_{1/2}$ - 2p$^{~2}$P$_{3/2}$ & 7.63& 3.71 & Z94 & & 2p$^{4~3}$P$_0$ - 2p$^{4~3}$P$_2$& 0.4568& 0.0524& \\
Mg IV&  2p$^{5~2}$P$_{3/2}$ -  2p$^{5~2}$P$_{1/2}$& 4.49& 0.36& S94& Ar XIV&  2p$^{~2}$P$_{1/2}$ - 2p$^{~2}$P$_{3/2}$ & 0.4414& 0.78 & Z94\\
Mg V& 2p$^{4~3}$P$_2$ - 2p$^{4~3}$P$_1$& 5.61& 0.85& A16& Fe I & 4s$^2$3d$^{6~5}$D$_4$ -  4s$^2$3d$^{6~5}$D$_3$& 24.04& 0.02& P97\\
&  2p$^{4~3}$P$_2$ - 2p$^{4~3}$P$_0$& 3.965& 0.22&  &  & 4s$^2$3d$^{6~5}$D$_4$ - 4s$^2$ 3d$^{6~5}$D$_2$& 14.20& 0.043& \\
Mg VII& 2p$^{2~3}$P$_0$ - 2p$^{2~3}$P$_1$& 8.873& 0.24& L94 & & 4s$^2$3d$^{6~5}$D$_4$ - 4s$^2$ 3d$^{6~5}$D$_1$& 11.26& 0.0001& \\
& 2p$^{2~3}$P$_0$ - 2p$^{2~3}$P$_2$& 3.403& 0.18& & Fe II& 3d$^6$($^5$D)4s a$^6$D$_{9/2}$ - 3d$^6$($^5$D)4s a$^6$D$_{7/2}$& 25.98& 6.27& Z95 \\
Mg VIII& 2p$^{~2}$P$_{1/2}$ - 2p$^{~2}$P$_{3/2}$ & 3.027& 0.85 & Z94 & & 3d$^6$($^5$D)4s a$^6$D$_{7/2}$ - 3d$^6$($^5$D)4s a$^6$D$_{5/2}$& 35.35& 1.85&\\
Si I& 3p$^{2~3}$P$_0$ - 3p$^{2~3}$P$_1$& 129.70& 0.2& H89& & 3d$^6$($^5$D)4s a$^6$D$_{5/2}$ - 3d$^6$($^5$D)4s a$^6$D$_{3/2}$& 51.23& 0.86&\\
& 3p$^{2~3}$P$_0$ - 3p$^{2~3}$P$_2$& 44.82& 0.2& & & 3d$^6$($^5$D)4s a$^6$D$_{3/2}$ - 3d$^6$($^5$D)4s a$^6$D$_{1/2}$& 87.68& 0.36&\\
Si II& 3p$^{~2}$P$_{1/2}$ - 3p$^{~2}$P$_{3/2}$ & 34.82& 5.23 & A14 & &3d$^7$ a$^4$F$_{9/2}$ -  3d$^7$ a$^4$F$_{7/2}$& 17.94& 1.70& \\
Si VI& 2p$^{5~2}$P$_{3/2}$ -  2p$^{5~2}$P$_{1/2}$& 1.965& 0.31& S94& Fe III& 3d$^{6~5}$D$_4$ -  3d$^{6~5}$D$_3$& 22.94& 2.85& Z96\\
Si VII& 2p$^{4~3}$P$_2$ - 2p$^{4~3}$P$_1$& 2.481& 0.31& Z96 & & 3d$^{6~5}$D$_4$ -  3d$^{6~5}$D$_2$& 13.53& 0.84& \\
&  2p$^{4~3}$P$_2$ - 2p$^{4~3}$P$_0$& 1.795& 0.066& & & 3d$^{6~5}$D$_4$ -  3d$^{6~5}$D$_1$& 10.73& 0.42& \\
Si IX& 2p$^{2~3}$P$_0$ - 2p$^{2~3}$P$_1$& 3.917& 0.33& L94 & & 3d$^{6~5}$D$_4$ -  3d$^{6~5}$D$_0$& 9.734& 0.13& \\
& 2p$^{4~3}$P$_2$ - 2p$^{4~3}$P$_2$& 1.558& 0.25& & Fe V& 3d$^{4~5}$D$_0$ -  3d$^{4~5}$D$_1$& 70.37& 0.861& B95\\
Si X&  2p$^{~2}$P$_{1/2}$ - 2p$^{~2}$P$_{3/2}$ & 1.432& 1.14 & Z94& & 3d$^{4~5}$D$_0$ -  3d$^{4~5}$D$_2$& 25.92& 0.461& \\
S I& 3p$^{4~3}$P$_2$ - 3p$^{4~3}$P$_1$& 129.70& 0.2& H89& & 3d$^{4~5}$D$_0$ -  3d$^{4~5}$D$_2$& 12.45& 0.392& \\
& 3p$^{4~3}$P$_2$ - 3p$^{4~3}$P$_0$& 44.82& 0.2& & & 3d$^{4~5}$D$_0$ -  3d$^{4~5}$D$_3$& 7.795& 0.338& \\
S III& 3p$^{2~3}$P$_0$ - 3p$^{2~3}$P$_1$& 33.48& 1.70& G95& Fe VI& 3d$^{3~4}$F$_{3/2}$ -  3d$^{3~4}$F$_{5/2}$& 19.56& 2.53& C99\\
& 3p$^{2~3}$P$_0$ - 3p$^{2~3}$P$_2$& 18.713& 0.81& & & 3d$^{3~4}$F$_{3/2}$ -  3d$^{3~5}$F$_{7/2}$& 8.415& 1.54& \\
S IV& 3p$^{~2}$P$_{1/2}$ - 3p$^{~2}$P$_{3/2}$ & 10.511& 6.89 & S99 & & 3d$^{3~4}$F$_{3/2}$ -  3d$^{3~4}$F$_{9/2}$& 0.4999& 1.38& \\
S VIII& 2p$^{5~2}$P$_{3/2}$ -  2p$^{5~2}$P$_{1/2}$& 0.989& 0.19& S94& Fe VII&  3d$^{2~3}$F$_{2}$ -  3d$^{2~3}$F$_{3}$& 9.51& 2.58& T14\\
S IX& 2p$^{4~3}$P$_2$ - 2p$^{4~3}$P$_1$& 1.255& 0.89& B94 &  &  3d$^{2~3}$F$_{2}$ -  3d$^{2~3}$F$_{3}$& 4.289& 1.53& \\
&  2p$^{4~3}$P$_2$ - 2p$^{4~3}$P$_0$& 0.041& 0.26& & Fe VIII& 3d$^{~2}$D$_{3/2}$ -  3d$^{~2}$D$_{5/2}$& 5.447& 2.9& D14\\
S XI&  2p$^{2~3}$P$_0$ - 2p$^{2~3}$P$_1$& 1.927& 0.088& L94& Fe X &2p$^{5~2}$P$_{3/2}$ -  2p$^{5~2}$P$_{1/2}$& 0.6376& 3.38& P95 \\
& 2p$^{4~3}$P$_2$ - 2p$^{4~3}$P$_2$& 0.808& 0.046& \\
S XII& 2p$^{~2}$P$_{1/2}$ - 2p$^{~2}$P$_{3/2}$ & 0.763& 0.12 & Z94\\
\hline
\end{tabular}
\end{center}
\tablecomments{K95 \citet{bhatia95}; L94 \citet{lennon94}; A08 \citet{aggarwal08}; W17 \citet{wang17}; Z94 \citet{zhang94}; H89 \citet{hollenbach89}; B94 \citet{butler94}; S99 \citet{saraph99}; S94 \citet{saraph94}; A16 \citet{aggarwal16}; P95 \citet{pelan95};
A14 \citet{aggarwal14}; G95 \citet{galavis95}; P97 \citet{pelan97}; S96 \citet{saraph96}; Z95 \citet{zhang95}; Z96 \citet{zhang96}; B95 \citet{berrington95}; C99 \citet{chen99}; T14 \citet{tayal14}; D14 \citet{delzanna14} }
\end{table}

\end{document}